\documentclass[]{aa}
\pdfoutput=1

\usepackage{graphicx}
\usepackage{txfonts}

\usepackage[backref,breaklinks,colorlinks=true,citecolor=blue]{hyperref}

\begin{document}

\title{Direct mapping of the temperature and velocity gradients in discs}
  \subtitle{Imaging the vertical CO snow line around IM Lupi}
  \titlerunning{Direct mapping of the temperature and velocity gradients in discs}

\author{C. Pinte
  \inst{1,2,3}
  \and
  F.~M\'enard\inst{1}
  \and
  G.~Duch\^ene\inst{4,1}
  \and
  T.~Hill\inst{5}
  \and
  W.R.F.~Dent\inst{5}
  \and
  P.~Woitke\inst{6}
  \and
  S.~Maret\inst{1}
  \and
  G.~van~der~Plas\inst{1}
  \and
  A.~Hales\inst{5,7}
  \and
  I.~Kamp\inst{8}
  \and
  W.F.~Thi\inst{9}
  \and
  I.~de~Gregorio-Monsalvo\inst{5}
  \and
  C.~Rab\inst{8}
  \and
  S.P.~Quanz\inst{10}
  \and
  H.~Avenhaus\inst{11}
  \and
  A.~Carmona\inst{12}
  \and
  S.~Casassus\inst{11}
}

\institute{Univ. Grenoble Alpes, CNRS, IPAG, F-38000 Grenoble, France\\
  \email{christophe.pinte@univ-grenoble-alpes.fr}
  \and
  UMI-FCA, CNRS/INSU, France (UMI 3386), and Dept. de
  Astronom\'{\i}a, Universidad de Chile, Santiago, Chile.
  \and
  Monash Centre for Astrophysics (MoCA) and School of Physics and Astronomy,
  Monash University, Clayton Vic 3800, Australia
  \and
  Astronomy Department, University of California, Berkeley
  CA 94720-3411, USA
  \and
  Atacama Large Millimeter / Submillimeter Array, Joint ALMA Observatory, Alonso de C\'ordova 3107, Vitacura 763-0355, Santiago, Chile
  \and
  Centre for Exoplanet Science, SUPA, School of Physics and Astronomy,
  University of St Andrews, North Haugh, St Andrews, Fife, KY16 9SS, UK
  \and
  National Radio Astronomy Observatory, 520 Edgemont Road, Charlottesville, Virginia, 22903-2475, United States
  \and
  Kapteyn Astronomical Institute, University of Groningen, Postbus 800, 9700 AV
  Groningen, The Netherlands
  \and
  Max Planck Institute for Extraterrestrial Physics, Giessenbachstrasse, 85741 Garching, Germany
  \and
  Institute for Astronomy, ETH Zurich, Wolfgang-Pauli-Strasse
27, 8093 Zurich, Switzerland
  \and
  Departamento de Astronomi\'ia, Universidad de Chile, Casilla 36-D, Santiago,
  Chile
  \and
  IRAP, Universit\'e de Toulouse, CNRS, UPS, Toulouse, France
}

\date{}

\abstract{Accurate measurements of the physical structure of protoplanetary discs are
  critical inputs for planet formation models. These constraints are traditionally
  established via complex modelling of continuum and line
  observations. Instead, we
  present an empirical framework to locate the CO isotopologue emitting
  surfaces from high spectral and spatial resolution ALMA observations. We apply this
  framework to the disc surrounding IM Lupi, where
  we report the first direct, {i.e.} model independent, measurements of the
  radial and vertical gradients of temperature and velocity in a protoplanetary disc.
  The measured disc structure is consistent with an irradiated
  self-similar disc structure, where the temperature increases and the velocity decreases
  towards the disc surface.
  We also directly map the vertical CO snow line, which is located at about one
  gas scale height at radii
  between 150 and 300\,au, with a CO freeze-out temperature of $21\pm2$\,K.
  In the outer disc ($> 300$\,au), where the gas surface density transitions from a power law to an
  exponential taper, the velocity rotation field becomes significantly sub-Keplerian,
  in agreement with the expected steeper pressure gradient.
  The sub-Keplerian velocities should result in a very efficient inward
  migration of
  large dust grains, explaining the lack of millimetre continuum emission
  outside of
  300\,au.  The sub-Keplerian motions may also be the signature of the base
    of an externally irradiated photo-evaporative wind. In the same outer
    region, the measured CO temperature above the snow line
  decreases  to $\approx$ 15\,K because of the reduced gas density, which
  can result in a lower CO freeze-out temperature, photo-desorption, or deviations
  from local thermodynamic equilibrium.
}

\maketitle

\section{Introduction}

\begin{figure*}
  \includegraphics[width=0.45\hsize]{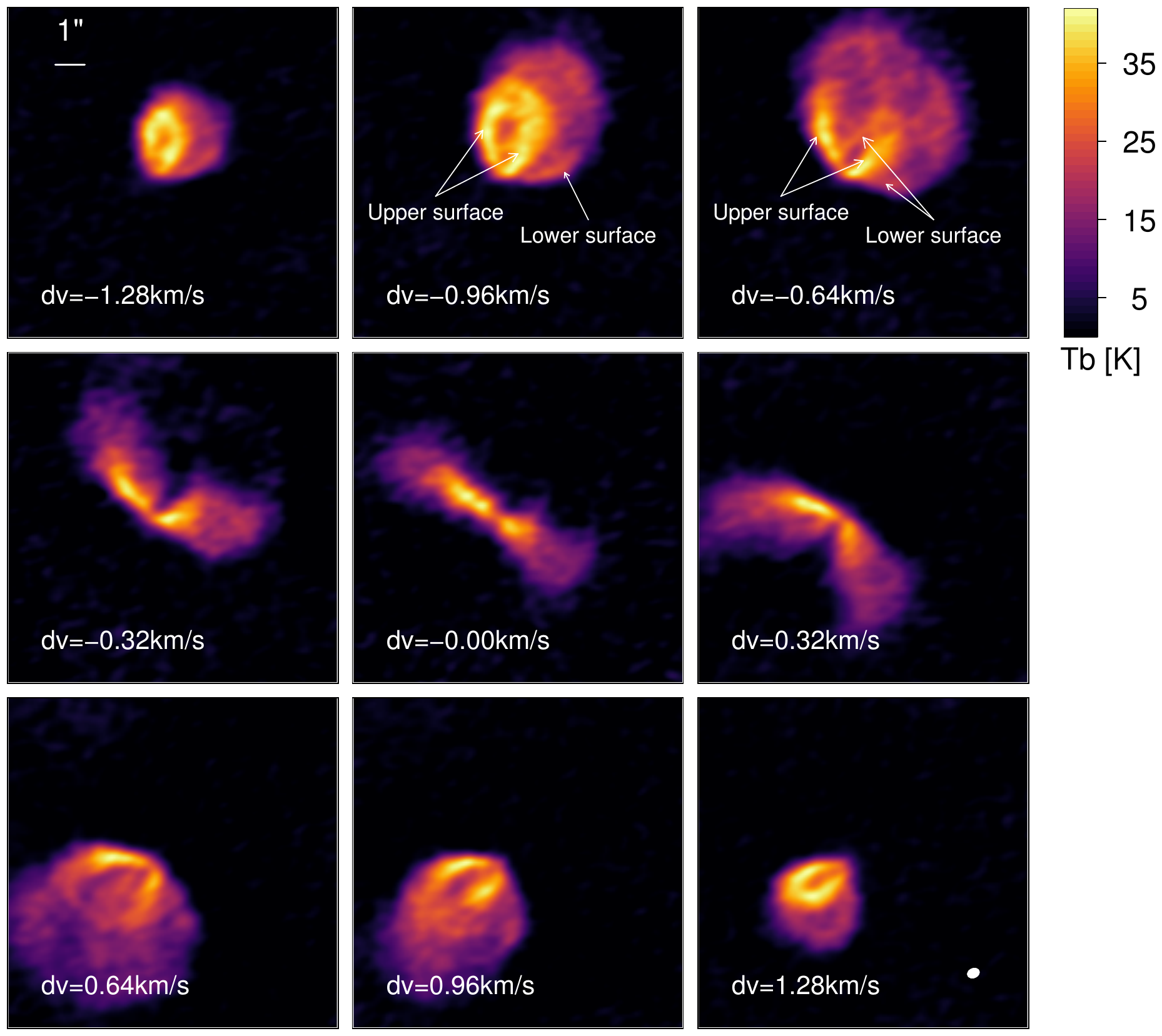}
  \hfill
   \includegraphics[width=0.45\hsize]{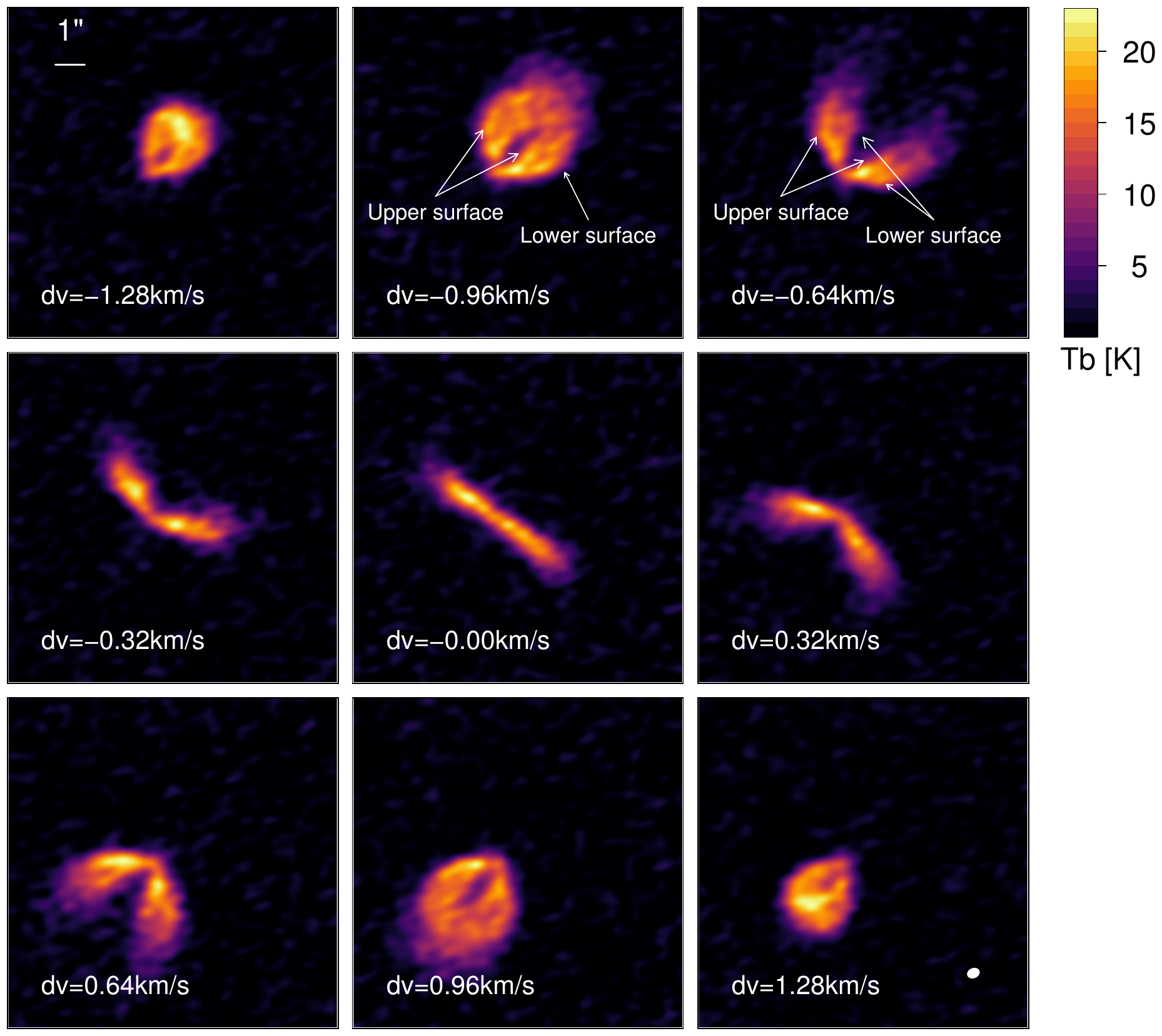}
  \caption{Channel maps of $^{12}$CO (\emph{left}) and $^{13}$CO (\emph{right})
    displayed with a channel width of 80\,m/s. The data is not continuum
    subtracted. The dv is relative to the systemic velocity of
    4.5\,km.s$^{-1}$. The beam is indicated in the bottom right corner.\label{fig:12CO}}
\end{figure*}

\begin{table*}
  \centering
  \begin{tabular}{lccc}
    \hline
    \hline
    Wavelength or line & Disc-integrated flux & Beam & Image or moment 0 RMS (1\,$\sigma$)\\
    \hline
    1.3\,mm cont. & 225 $\pm$ 23\,mJy & 0.36'' $\times$        0.55'' at
    -80$^\circ$ & 0.05\,mJy/beam\\
    $^{12}$CO (2-1) &  24.1 $\pm$    2.4 Jy.km/s & 0.36'' $\times$  0.56'' at
    -78$^\circ$ & 3.5\,mJy.km/s\\
    $^{13}$CO (2-1) & 7.9 $\pm$     0.8 Jy.km/s & 0.38'' $\times$  0.58'' at
    -79$^\circ$ & 4.2\,mJy.km/s\\
    C$^{18}$O (2-1) &  1.3 $\pm$   0.2 Jy.km/s & 0.38'' $\times$  0.58'' at
    -78$^\circ$ & 3.9\,mJy.km/s\\
    \hline
  \end{tabular}
  \caption{Summary of observations. The continuum flux was measured over
      the area where the signal is greater than 3 times the map RMS (estimated far from
      the source). The line flux was measured by integrating the flux where the
      signal is greater than 3 times the per channel RMS, estimated from the scatter
      in  line-free channels. \label{tab:fluxes}}
\end{table*}

The initial stages of planet formation are the results of a complex interplay
between several physical mechanisms
which all depend on the disc density, temperature,
chemical abundances, and  velocity structure \citep{Testi14PPVI}.
A critical step is the growth of  submicron-sized grains,
originating from the parent cloud, into large planetesimals
within the few million years of a protoplanetary disc lifetime.
Dust grains are believed to grow by collision and sticking  \citep{Blum08,Dominik06PPV}.
In parallel, they settle towards the
disc midplane and migrate inwards as a result of the conjugate actions
of  stellar gravity, aerodynamical gas drag, and radial pressure gradient
\citep[{e.g.}][]{Weidenschilling77}.
The radial concentration of dust grains depends strongly on the
sub-Keplerian gas velocity, pressure gradients
\citep[{e.g.}][]{Pinilla12}, and disc thermal structure
\citep{Laibe12,Pinte14}. In particular, molecular snow lines -- the
two-dimensional surface where abundant volatiles condensate
onto dust grains -- result in increased solid particle stickiness and
dust-to-gas ratio, promoting particle growth \citep[{e.g.}][]{Ciesla06,Ros13}.

These dynamical processes remain hard to constrain by existing observations,
however. Rotational molecular lines
are a powerful tool that can be used to probe
the disc structure, but constraints
rely in most cases on complex model fitting or on indirect methods.
For instance, detailed radiative transfer modelling of multiple CO lines can provide an
estimate of the vertical temperature structure of
the disc \citep{Dartois03}, indicate freeze-out of the gas phase molecules
onto dust grains in the cold midplane \citep[{e.g.}][]{Qi11,Zhang17}, and suggest a very low level
of turbulence \citep[{e.g.}][]{Flaherty15}. Indirect mapping via chemical imaging
(\citealp{Qi13} with  N$_2$H$^+$ and \citealp{Mathews13} with DCO$^+$) has also been used to estimate the location of the CO snow
line.  However,  \cite{Aikawa15}, \cite{van't_Hoff17}, and \cite{Huang17}
  have shown that
  these species are not robust CO snow line tracers, and that detailed chemical
modelling is required to infer the midplane CO snow line location from N$_2$H$^+$ or
DCO$^+$ observations.

The combination of high spatial and spectral resolution and sensitivity
offered by ALMA opens new avenues to directly map the disc thermal and
kinematic structure by resolving the gas disc both radially and vertically.
\cite{Dutrey17} recently introduced a method to map the thermal and density gas structure of
 discs at close to edge-on inclinations.
Discs at intermediate inclinations are also ideal targets as the
Keplerian velocities spatially separate the emitting regions,
eliminating line of sight confusion \citep[{e.g.}][]{deGregorio13,Rosenfeld13}.

IM Lupi is a M0V T Tauri star (distance of
$161\pm10$\,pc, \citealp{Gaia16}) surrounded by a large and massive disc with
an inclination of $48^\circ\pm3^\circ$ \citep{Pinte08,Cleeves16}.
The disc is detected in rotational CO emission up to a radius of 970\,au, in the millimetre continuum up to 310\,au \citep{Panic09,Cleeves16},
and in scattered light up to 320\,au with a well-defined disc morphology \citep[][rescaled
to the updated distance]{Pinte08}, but with an extended component up to
720\,au in radius.
We present here a framework to {directly} measure the altitude, velocity, and
temperature of the CO layers from new high spectral (0.05\,km/s) and
intermediate spatial (0.4'') resolution
ALMA observations of the disc surrounding IM~Lupi.

\section{Observations and data reduction}

IM Lupi was observed with ALMA in band 6 on the night from  9
to
10  June 2015 with a total
on-source time of 37.4\,min (Program ID 2013.1.00798.S). The array was configured with 37
antennas and baselines ranging from 21.4 to 784\,m. Titan was used as a flux
calibrator and the quasars J1517-2422
and J1610-3958 were used as bandpass and phase calibrators.
Two of the four spectral
windows of the correlator were configured in Time Division Mode (TDM)  centred
at 218.9\,GHz and 231.7\,GHz, each with 1.875\,GHz usable bandwidth.
The other two spectral
windows were configured in Frequency Division Mode (FDM) to target the $^{12}$CO
J=2-1, $^{13}$CO J=2-1, and C$^{18}$O J=2-1 transitions, with a channel spacing of
30.5\,kHz and a channel averaging of 2.
Because of the Hanning
smoothing used to reduce spectral ringing, the actual spectral resolution is 35\,kHz,
corresponding to 45 -- 48\,m/s depending on the line.
The median partial water vapour was 0.93\,mm.
The data sets were calibrated using the Common Astronomy
Software Applications pipeline (CASA, \citealp{McMullin07}, version
4.6.0).
We performed three successive rounds of phase self-calibration, the last
with solutions calculated for each individual integration (6s),
followed by a phase and amplitude self-calibration.
The continuum
self-calibration solutions were applied to the CO lines.
A CLEAN mask was manually defined for the continuum image and each channel
map. Channel maps
were produced both with and without  continuum subtraction (using the CASA task
\emph{uvcontsub}).  We estimate the flux calibration uncertainty to 10\,\%.

The disc is spatially resolved in all transitions and channel maps show the typical butterfly pattern
of discs in Keplerian rotation.
The measured fluxes are presented in Table~\ref{tab:fluxes} and the
representative channel maps for $^{12}$CO and $^{13}$CO in Fig.~\ref{fig:12CO}.
The  C$^{18}$O channel maps and continuum image are shown in
Figs.~\ref{fig:C18O} and \ref{fig:continuum_image}, respectively.
Our data and results are consistent with those of
\cite{Cleeves16}. We focus in the following on the spatial distribution of the emission in
individual channel maps.

\section{Reconstructing the altitude, velocity, and temperature of the CO emitting layers}

\begin{figure}
  \centering
\includegraphics[width=0.9\hsize]{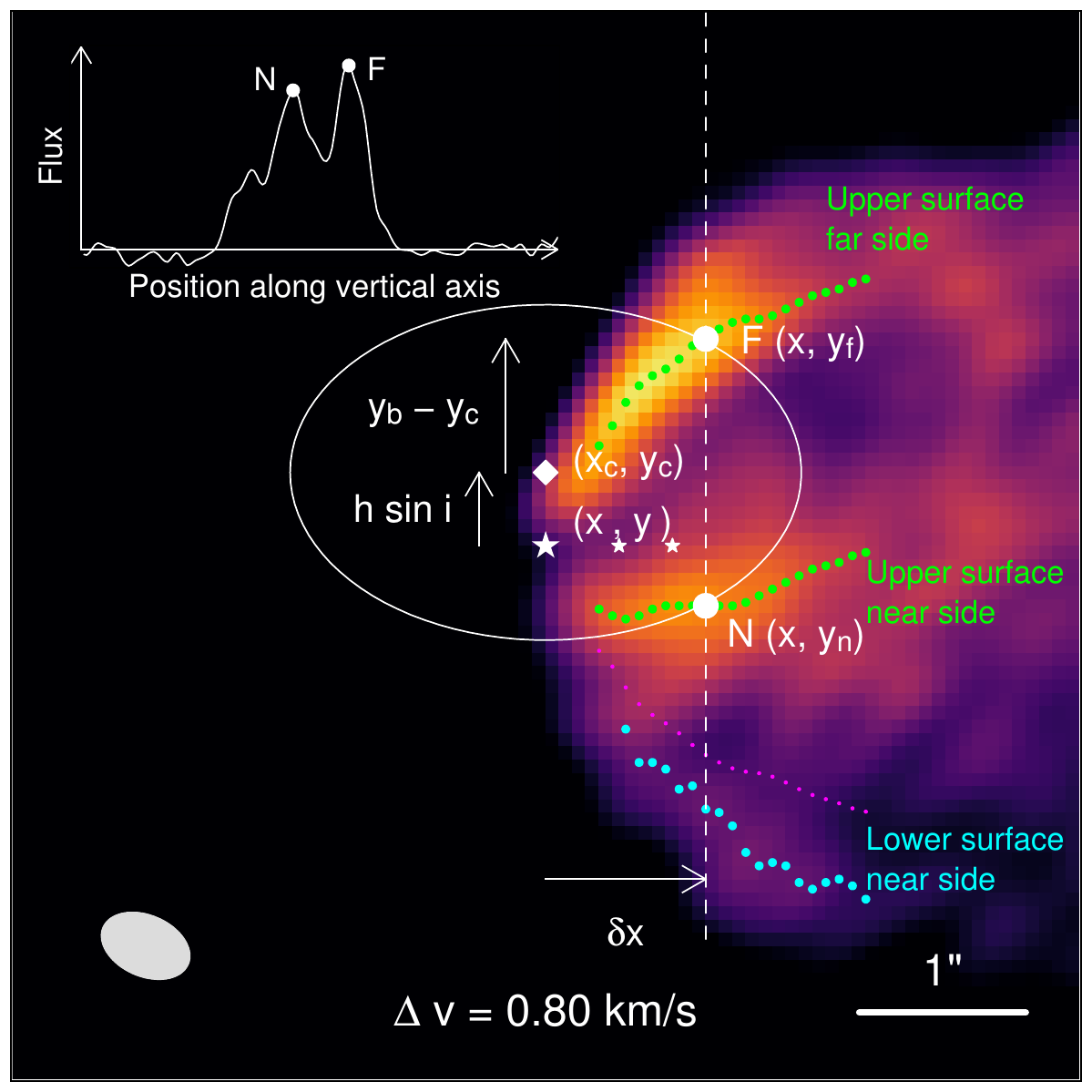}
  \caption{Single $^{12}$CO channel with a schematic
    of the
  quantities we measure. The disc has been rotated so that the
  semi-major axis is horizontal.
  The position of the central object is marked by a star.
  For a given offset $\delta x$ along the disc major axis, two maxima N and F on
  the same vertical are located on the near and far sides of the disc's upper surface.
  These two points are on the same inclined circular orbit (white ellipse), whose centre is
  marked by a diamond.
  Indicated are the geometrical parameters from which the cylindrical
  radius, height, and velocity of the points F and B can be measured.
  By repeating the procedure for each $\delta x$ (green points, sampled
    every 0.08'') and each
  velocity channel, a full mapping of the CO upper surface is performed.
  The cyan points mark the location of the lower surface (near side), while the magenta points
  represent the symmetric of the far upper surface relative to the disc major axis.
\label{fig:one_channel}}
\end{figure}

In a given channel map, the line emission is concentrated along the isovelocity curve, {i.e.} the set of points where the
projected velocity of the emitting surface is constant.
As a result, the emission originating from  the
upper and lower surfaces ({i.e.} above and below the midplane as seen
from the observer), and from the near and far
sides of these surfaces, is well separated (Figs.~\ref{fig:one_channel} and \ref{fig:12CO}).
By locating the emission in each channel map,
we can directly reconstruct, via simple
geometrical arguments, the position and velocity of each of the CO layers.

\subsection{Altitude}
\label{sec:h_CO}

We consider the coordinate system where the $x$-axis is aligned with the disc major axis, and we denote ($x_\star$, $y_\star$) the
position of the star which was determined by the peak of the continuum map (Fig.~\ref{fig:one_channel}).
Because the gas is vertically pressure supported, we assume that, at any point
in the disc, the gas is rotating on a
circular orbit parallel to the disc midplane, {i.e.} we assume that the bulk
of the motion is described by Keplerian rotation and neglect for this
analysis any radial or vertical gas circulation.  The radius and
altitude of this orbit are denoted by $r$ and $h$, respectively.

For a given offset $\delta x = x - x_\star$ in the image plane, a vertical line
will intersect the isovelocity curve of the upper disc surface\footnote{The
same formalism can be used for the lower surface, but for reasons we detail in
Sect. \ref{sec:results}, it cannot be used here and we only describe the
formalism for the upper surface.} at two points (for small enough $x$), which belong to the same orbit, {i.e.} at the same
distance from the star and at the same altitude.
Due to line broadening, and to finite spatial and
spectral resolution, the emission is not located exactly on the
isovelocity curve, but forms a narrow band around it. We estimate the position of the
emission by its maximum, as illustrated in Fig.~\ref{fig:one_channel}.
If we denote $y_n$ and $y_f$ the ordinates of the two
points on the near and far side of the disc, the coordinates of the centre of
the projected circular orbit passing through those two points are ($x_c$,
$y_c$), where $x_c = x_\star$ and $y_c = (y_f + y_n)/2$.
Any point on this ellipse fulfils
$(x-x_\star)^2 + ((y-y_c)/\cos i)^2 = r^2$,
where the orbital radius $r$ is the length of the semi-major axis of the
ellipse
and $i$ the disc inclination.

\begin{figure}
  \centering
  \includegraphics[width=\hsize]{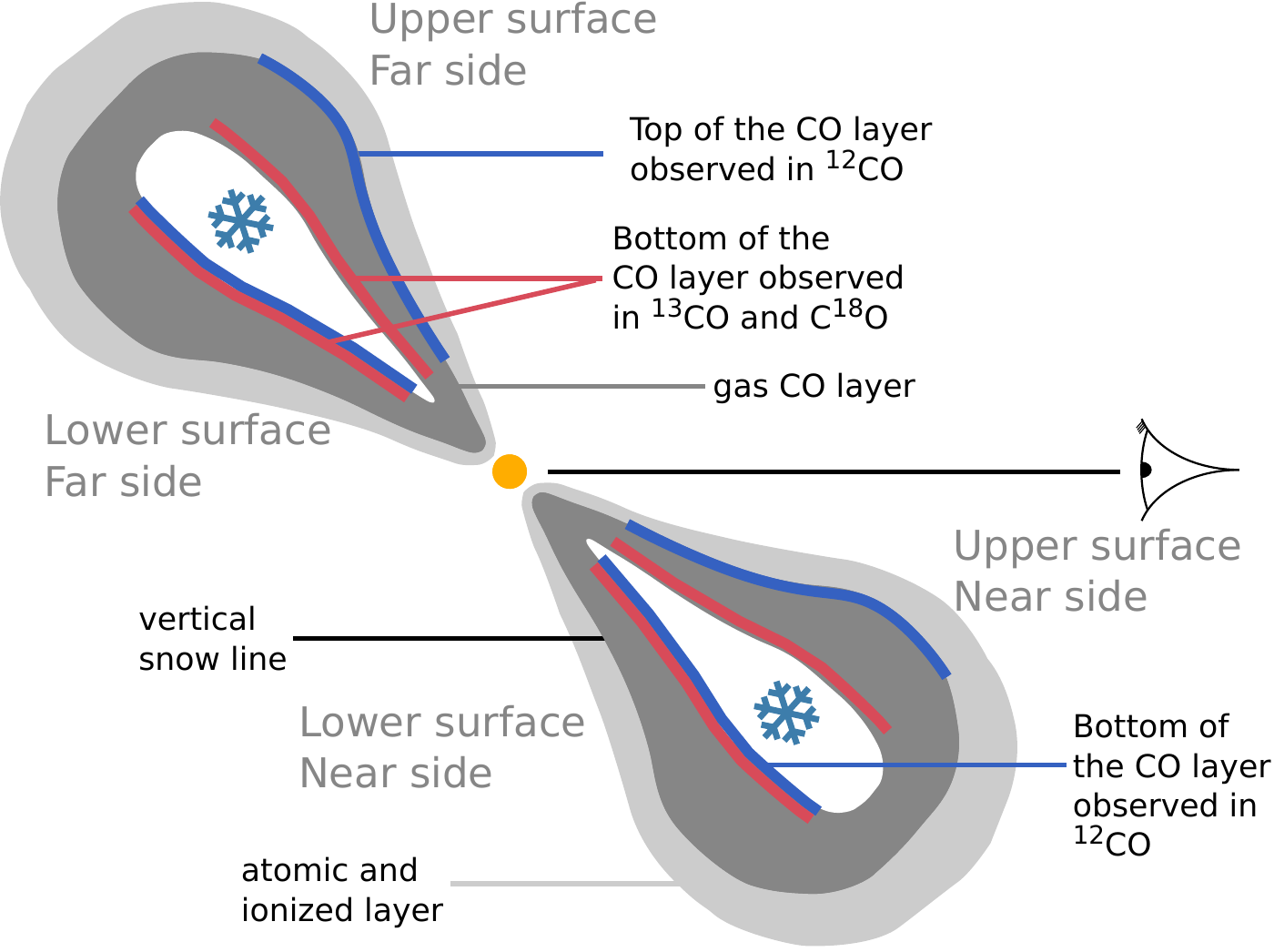}
  \caption{Schematic view of the various layers observed in the CO lines.\label{fig:schema}}
\end{figure}

\begin{figure*}
\includegraphics[width=\hsize]{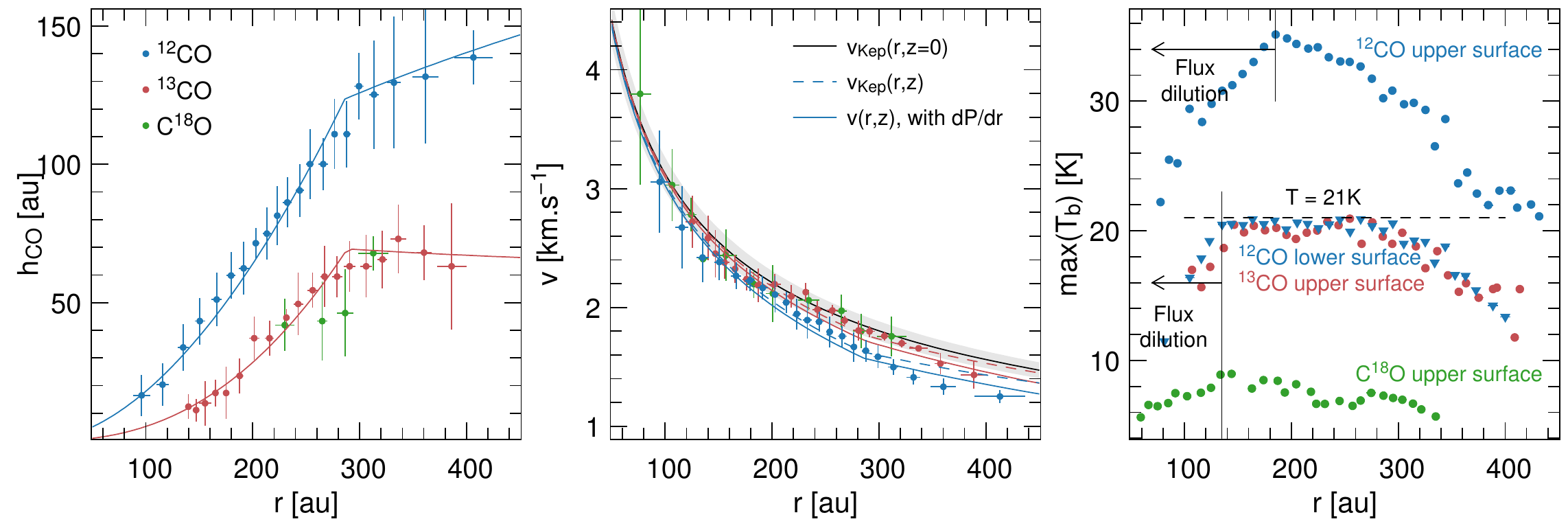}
\caption{
  Measured altitude, velocity, and brightness temperature of the CO
  layers as a function of radius: $^{12}$CO (blue), $^{13}$CO (red), and C$^{18}$O
  (green).
  Data points were extracted for all channel maps and binned according
  to distance. The error bars represent the dispersion within the bins added
  quadratically to the uncertainty resulting from inclination ($48\pm3^\circ$).
  \emph{Left:} Solid lines display a
  broken power-law fit (below and above $r = 300$\,au) for the $^{12}$CO and
  $^{13}$CO. \emph{Centre:} Black solid line represents the Keplerian velocity
  for a 1\,M$_\odot$ central star,
  derived from the C$^{18}$O position--velocity diagram (Appendix
  \ref{app:C18O}). The shaded area
  represents an uncertainty of 0.1\,M$_\odot$. The red and blue dashed
  lines show the expected Keplerian velocities taking into account the
  altitude of the  $^{12}$CO and $^{13}$CO layers. The red and blue solid lines display
  the velocity  of the  $^{12}$CO and $^{13}$CO layers taking into account the altitude and
  the pressure gradient term, as derived from our MCFOST model. \emph{Right:} Circles and
  triangles
  represent the temperature measured on the upper and lower surfaces,
  respectively. The brightness temperatures are similar on the
upper and lower surfaces for $^{13}$CO and C$^{18}$O; for clarity, only  the upper surface
is plotted.
  \label{fig:h_CO}}
\end{figure*}

By using the point on the far side of the disc surface, we can for instance derive
\begin{equation}
  r = \sqrt{(x-x_\star)^2 + \left(\frac{y_f-y_c}{\cos i}\right)^2}.
\end{equation}

The altitude $h$ of the orbit is derived by noting that the offset between the centre of the ellipse and
the star is simply $h \sin i$:
\begin{equation}
h = \frac{y_c - y_\star}{\sin i}.
\end{equation}

\subsection{Gas velocity}

In a given velocity channel, the projected radial velocity
$\varv_\mathrm{obs}$ is known and can be expressed as
$\varv_\mathrm{obs} = \varv_\mathrm{syst} + \varv\,\cos \theta\,\sin i$,
where $\varv$ is the actual gas velocity around the central star,  $\varv_\mathrm{syst}$ the systemic
velocity, and $\theta$ is the true longitude. By
noting that $\cos \theta = (x-x_\star) / r $, we obtain the
actual gas velocity at each point:
\begin{equation}
\varv  = (\varv_\mathrm{obs} -  \varv_\mathrm{syst})\,\frac{r}{(x - x_\star)\,\sin i}.
\end{equation}

\subsection{Gas kinetic temperature}

For optically thick emission, and as long as the emission
fills the beam, the observed brightness
temperature $T_\mathrm{b}$
is very close to the excitation temperature  $T_\mathrm{ex}$,
$T_\mathrm{b} = T_\mathrm{ex} (1 - \exp(-\tau))$.
Low~J CO lines are close to local thermodynamic equilibrium,
with an excitation almost equal to the gas kinetic temperature $T_\mathrm{kin}$
\citep[{e.g.}][]{Pavlyuchenkov07}.
For optically thick lines,
the peak intensity then provides an accurate estimate of the gas temperature.
This is the case for the $^{12}$CO and
  $^{13}$CO lines (see discussion in section~\ref{sec:results}), but
 the C$^{18}$O emission is partly optically thin and the brightness temperature is lower
than the actual excitation temperature.
To reconstruct the disc temperature profile, we measured the brightness
temperature at all points extracted above.
We also extracted  the brightness temperature of points on the disc
lower surface in the same way (Fig.~\ref{fig:one_channel}).

\subsection{Reconstructing the full spatial, kinematic, and thermal structure of the CO layers}

As described above, by locating the position of the maxima in each isovelocity curve and for
each $x$ in a given channel map (where we sampled $x$ every 0.08'', {i.e.}
  $\approx$ 1/4 of the beam),
we can directly measure $r$ and the corresponding
elevation $h$ and velocity $\varv$, and we can estimate $T_\mathrm{gas}$.

Each channel probes a range of distances from the star depending on the velocity.
By repeating this procedure for each channel, we can reconstruct the
complete 2D velocity and temperature distribution of the CO emitting layers, as shown
in Fig.~\ref{fig:h_CO}.

\subsection{Caveats}

  The central channels of $^{12}$CO (within $\pm$0.7\,km/s from the systemic velocity) suffer from strong foreground extinction
  \citep{Panic09,Cleeves16}. They were
  excluded from the analysis for the temperature estimate. The morphology of
  the channel maps is not affected, however, and we included them to measure the
  altitude and velocity.

The method presented in this section
can only be applied in regions where the emission
 originates from a vertically thin layer, so that an
  altitude $h$ can be defined, {i.e.} a clear maximum can be detected in
  the channel map. This is only possible when the projected width of the
  emitting layer is
  significantly smaller than the beam.
  This is the case
for instance when the molecular layer itself is a
narrow layer, or when the optically depth gradient is steep around the
$\tau=1$ surface and we can only see a
narrow layer.

These conditions are met for the CO layers of IM~Lupi up to a radius 450\,au,
  where we are able to
distinctly identify the location of the isovelocity curves,
 but are no longer valid far from the
star even though the emission is detected up to
$\approx$1\,000\,au.
As already noted by \cite{Cleeves16},
the CO emission at large scales ($> 450\,$au) becomes very diffuse, making it
impossible to locate the isovelocity curve and to estimate an altitude.
This is very likely the result of UV photo-desorption of the CO by external irradiation from the
interstellar radiation field  (see e.g. Fig.~\ref{fig:models} d),
allowing gas CO to exist even in the midplane and to emit from a vertically
extended region.

The presented framework also requires sufficient signal in a given channel to accurately locate the
emission. For $^{12}$CO we were able to use the intrinsic channel width, but for
$^{13}$CO and C$^{18}$O we needed to bin the data to channel widths of 80 and
320\,m/s, respectively. This introduces spatial smearing due to the disc Keplerian
velocity and reduces correspondingly the range over which we can accurately
measure the CO layer altitude\footnote{This mainly affects our
ability to measure $h$, which depends on the small distance between the star's
position and the centre of the gas orbit. The radius and velocity are less
affected as they depend on greater lengths in the channel map.}. For  $^{13}$CO and C$^{18}$O, this prevented a measurement of the
  altitude at radii smaller than $\approx\,150$\,au and $\approx\,220$\,au, respectively.

Continuum subtraction can significantly affect
the measured brightness temperature and apparent morphology of line emission when
the line is optically thick, and the line and
continuum intensities are comparable \citep{Boehler17}. At the centre of an
optically thick line, the dust continuum is absorbed by the molecule. As
the continuum is usually estimated from line-free
channels, this results
in oversubtraction of the continuum close to the line centre.
The authors conclude that
using non-continuum-subtracted maps is preferable in order to estimate the line
brightness temperature (see e.g. their Fig.~8).
This conclusion also holds for the framework we have presented, and using non-continuum-subtracted channel maps  will provide a more accurate estimate of the
brightness temperature and will prevent artificial spatial shift of the line
emission due to oversubtraction of the local continuum emission.

At the spatial resolution of our observations, the continuum subtraction has
no impact on the results as the continuum is optically thin in the regions where we
can isolate the various isovelocity curves, and the corresponding continuum
intensity is negligible compared to
the line intensity. As a sanity check, we performed our analysis on both the
continuum-subtracted and non-subtracted maps and did not find any significant
difference.
 At higher spatial resolution, however,  the effect is very likely to become
important as the continuum emission in our current central beam is
significant, producing a central `hole' in the continuum-subtracted $^{13}$CO and
C$^{18}$O maps \citep{Oberg15,Cleeves16}, which is absent in the non-continuum-subtracted maps (Fig.~\ref{fig:12CO} and \ref{fig:C18O}).

\section{Results and discussion}
\label{sec:results}

\subsection{ CO layers}

 As expected, we find that the
$^{12}$CO emitting layer lies higher in the disc than the $^{13}$CO, due
 to its much higher optical depth (Fig.~\ref{fig:h_CO}, left panel). This is
 readily visible in
 Fig.~\ref{fig:12CO}, where the two surfaces are geometrically much closer to one
  another for $^{13}$CO than $^{12}$CO.
While already suggested by
previous ALMA observations
\citep[{e.g.}][for HD~163296]{deGregorio13,Rosenfeld13},
this is the first {direct}, {model-independent} measurement of the
altitude of the CO layers.
In particular, we clearly see that CO surfaces are flaring and seem to
flatten out outside of $\approx 300\,$au. At a distance of 200\,au, the
$^{12}$CO emission originates from an altitude of $\approx$65\,au, while the
$^{13}$CO originates from $\approx$25\,au. This corresponds to about 2.5 and 1
hydrostatic scale height, respectively. Interestingly, these values are very
  similar to the numbers that \cite{Dartois03} and \cite{Dutrey17} have found
  for DM Tau and the Flying Saucer, respectively.

  A linear regression analysis indicates a power-law
exponent of 1.8$\pm$0.2 and 2.1$\pm$0.4 for the altitude of $^{12}$CO and
$^{13}$CO layers at radii smaller than 300\,au, respectively. It should be noted that these values
are much larger than the actual faring index of the disc pressure scale height
($\approx$ 1.15 - 1.2 in the various published disc models).
This is expected, however, as the shape of the emitting CO layer is
set by the irradiation from the star (responsible for both the photo-dissociation and
disc temperature). The emitting molecular layer should roughly follow a layer with a constant optical depth as seen from
the star, and not the actual disc scale height.  Similarly, large exponents
were measured around the Herbig star HD\,97048 for the PAH
emission \citep{Lagage06} and scattered light \citep{Ginski16} layers, which
are also set by the stellar irradiation.

The C$^{18}$O emission is more difficult to locate due to the lower signal-to-noise ratio and
required velocity binning, but it appears co-spatial with the $^{13}$CO
emission.
Since $^{13}$CO and C$^{18}$O have a typical optical depth ratio of about 8:1,
one expects the C$^{18}$O to be closer to the midplane than $^{13}$CO by the
same reasoning as for $^{13}$CO and $^{12}$CO. The fact that we do not see any
significant difference in altitude between $^{13}$CO and C$^{18}$O suggests that we are actually probing a physical
layer in the disc below which there is no significant CO emission, hence this layer marks the
vertical freeze-out line.

\begin{figure*}
  \centering
\includegraphics[width=\hsize]{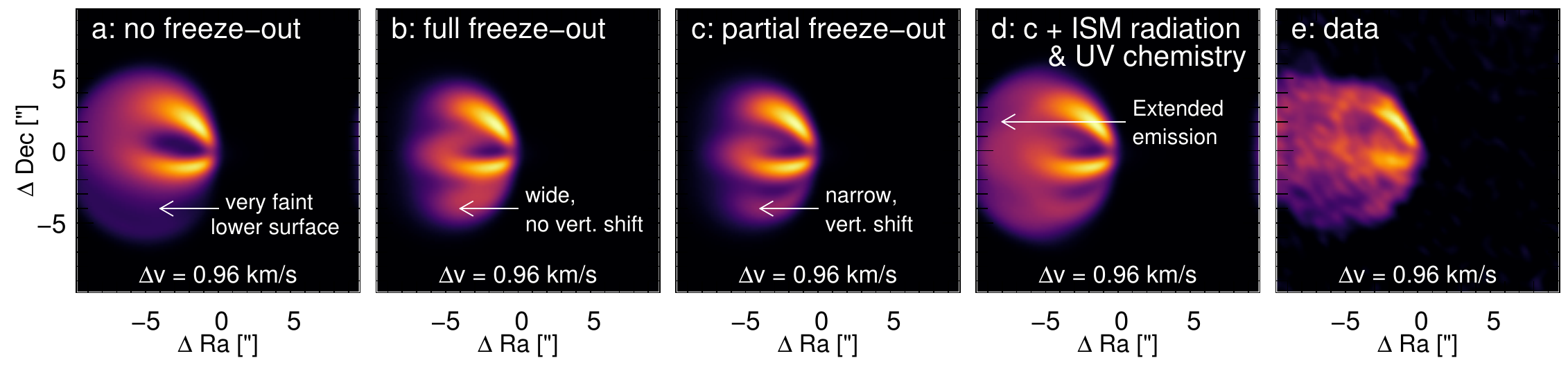}
\caption{Representative models of IM Lupi. The disc has been rotated so that
  the semi-major axis is horizontal. From left to right \emph{a}:
  without any freeze-out of CO; \emph{b}: with complete freeze-out of CO, where
  T$<21$\,K; \emph{c}: with depletion of gaseous CO by $10^{-4}$, where
  T$<21$\,K; \emph{d}: same as \emph{c} including the interstellar radiation
  field. The models are designed for  qualitative comparison only.
  They are convolved to the ALMA resolution, but not degraded to the same noise level as the observations.
  \label{fig:models}}
\end{figure*}

\subsection{ CO depleted midplane}

The disc lower surface ({i.e.} on the other side of the disc midplane)
is clearly visible, in particular for
$^{12}$CO, telling us that the CO gas abundance  is significantly reduced in the
disc midplane. To illustrate this, we generated a representative disc
model (Fig.~\ref{fig:models}) using the radiative transfer code MCFOST \citep{Pinte06,Pinte09}.
This model
qualitatively explains the observed morphological features,
but is not in any sense optimized and should only be considered for
general guidance. We also
produced a few additional models where we modified physical ingredients
(UV chemistry, CO condensation) to discuss their impact on the observed channel
maps (Fig.~\ref{fig:models}).  The  model description and parameters are presented in
Appendix~\ref{app:model_parameters}. In particular, we see that the CO needs
to be physically depleted in the disc midplane to be able to observe the lower disc
surface. The temperature gradient alone (Fig.~\ref{fig:models}, panel a) results in a strong
attenuation by the cold CO in the midplane, masking the lower surface.

Because the $^{12}$CO emission on the lower surface is seen through the disc, it
should originate from the
lowest boundary of the actual CO layer, {i.e.} exactly at the vertical CO snow
line.  It can potentially provide a more accurate estimate of the altitude of
the CO snow line, and of the CO layer
thickness by comparison with measurements from the upper surface.
We attempted to locate the $^{12}$CO lower surface using the same formalism as for the upper
surface. But, surprisingly, the lower surface appears shifted further away from the midplane
than the upper surface is (Fig.~\ref{fig:one_channel}), indicating an apparent
higher altitude, while one would expect the opposite (see Fig.~\ref{fig:schema}). The $^{13}$CO does
not show any shift on the other hand. The absence of shift is consistent with
the $^{13}$CO emission probing the actual physical
layer of CO. We interpret the apparent shift of the lower $^{12}$CO surface as
a radiative transfer effect.

If the CO is entirely depleted at temperatures below 20\,K, the lower
disc surface appears very wide and uniformly bright in the model, with no shift
relative to the upper surface (Fig.~\ref{fig:models}b).
To reproduce the observed morphology, we need a partial depletion of
$\approx 10^{-4}$ ({i.e.} an abundance of $\approx 10^{-8}$)
of CO where T\,$<$\,20\,K (Fig.~\ref{fig:models}c). The
  exact value is not well constrained and
  depends on other parameters, in particular the
  local turbulent velocity dispersion.
This small fraction of cold CO gas below
the snow line attenuates the centre of the emission from the lower
surface. Emission can only escape where the velocity shift is large enough,
{i.e.} slightly away from the isovelocity curve,
resulting in an apparent spatial shift of the lower surface. In our
simple model, this is reproduced by a constant depletion factor. In reality,
this means that the vertical snow line is not infinitely sharp, as expected, but with
a CO abundance that decreases progressively with altitude, {i.e.} over a few
K. It is potentially a powerful diagnostic to constrain the physics of the CO
condensation on grains, and the possible vertical `smearing' of the ice line
due to turbulent vertical mixing \citep{Aikawa07,Furuya14}.

\subsection{Vertical velocity gradient and sub-Keplerian rotation}

The different altitudes probed by the CO isotopologues also allow for the first
measurement of the vertical gradient on the rotational velocity of the
disc. Figure~\ref{fig:h_CO}  clearly shows that the $^{12}$CO is rotating slower
than the
$^{13}$CO and C$^{18}$O. This is expected as the radial projection of the
gravitational force
by the star per unit mass decreases with altitude:
\begin{equation}
  \frac{\varv^2}{r} = \frac{r}{\sqrt{r^2 + h^2}} \times  \frac{G\,M_\star}{r^2 + h^2}.
\end{equation}
Within 300\,au, the difference in the measured velocity fields between
$^{12}$CO and $^{13}$CO/C$^{18}$O is in good agreement with the
vertical dependence of the gravitational field (Fig.~\ref{fig:h_CO}, central panel).

At larger radii, the gas becomes clearly
sub-Keplerian.  This can be understood by including the pressure
gradient term \citep[{e.g.}][]{Rosenfeld13}:
\begin{equation}
  \frac{\varv^2}{r} = \frac{G M_\star r}{(r^2 + h^2)^{3/2}}
  + \frac{1}{\rho_\mathrm{gas}} \frac{\partial P}{\partial r}.
\end{equation}
With both the density and temperature profile generally decreasing with
distance, this term is negative, reducing the gas velocity compared to the
fiducial Keplerian velocity. This phenomenon is thought to play an important
role in the radial migration
and growth of solids in these discs \citep{Weidenschilling77,Barriere05,Brauer08}.

Interestingly, the radius at which we see significant sub-Keplerian velocities
coincides with the radius where the gas surface density transitions from a
power law to a tapered profile \citep{Panic09,Cleeves16}, {i.e.} where
the disc pressure gradient is
becoming larger and where the  velocities should deviate the most from Keplerian.
At a radius of 400\,au, the measured deviations from the altitude
  dependent Keplerian rotation
  are of the order of $0.15$\,km/s for both $^{13}$CO and $^{12}$CO.
This is
consistent with the expected pressure gradient as derived from the temperature
and density structure of our MCFOST model (Fig.~\ref{fig:h_CO}), which
  predicts a slightly smaller correction of $\approx 0.1$\,km/s at
  400\,au. This value depends on the local density and temperature
  gradients, which are not well constrained at these distances from the star,
  and the pressure gradient term might be able to fully explain  the observed deviations.
   We note that the deviations from Keplerian velocities start at
  slightly different radii for the two isotopologues, $\approx$300\,au for
  $^{12}$CO and $\approx$350\,au for
  $^{13}$CO. The difference is about one beam size and would need to be
  confirmed with higher spatial resolution observations.
  This behaviour is not captured by our model, suggesting that an additional
  process might be at play.

A possible explanation is that we could also be observing the
  base of a photo-evaporative wind. \cite{Haworth17} showed that the large-scale halo
of CO emission ($> 1\,000\,$au) described by \cite{Cleeves16} could be
interpreted as the result of a photo-evaporative wind created by irradiation
from the local (weak) external radiation field. The CO abundance in the flow
remains high enough to be comparable to the observed large-scale emission around IM
Lupi. Interestingly, as shown by \cite{Facchini16} and \cite{Haworth16b}, such
a photo-evaporative wind
can also result in a sub-Keplerian rotation field (see e.g. Fig.~10 of
\citealp{Haworth16b}), between
the outer edge of the disc and the H-H$_2$ transition. For the 1D models of IM
Lupi presented by \cite{Haworth17}, the expected deviation from Keplerian
rotation is at the 0.1 - 0.2\,km/s level (Haworth, private communication) and
would also be
consistent with our observations. Such a photo-evaporative wind could also
explain why the deviations from Keplerian rotation seem to start closer in at
high altitude where the density is lower, allowing the external radiation to
penetrate further in. However, both \cite{Facchini16} and \cite{Haworth16b}
predict that the wind should also be
associated with a radially increasing gas temperature profile. This is clearly
not
the case for IM Lupi, and more detailed modelling ({e.g.} 2D) might be
necessary to test  the hypothesis of external photo-evaporation further.

\subsection{Two-dimensional temperature structure and vertical CO snow line}

\begin{figure}
  \centering
  \includegraphics[width=\hsize]{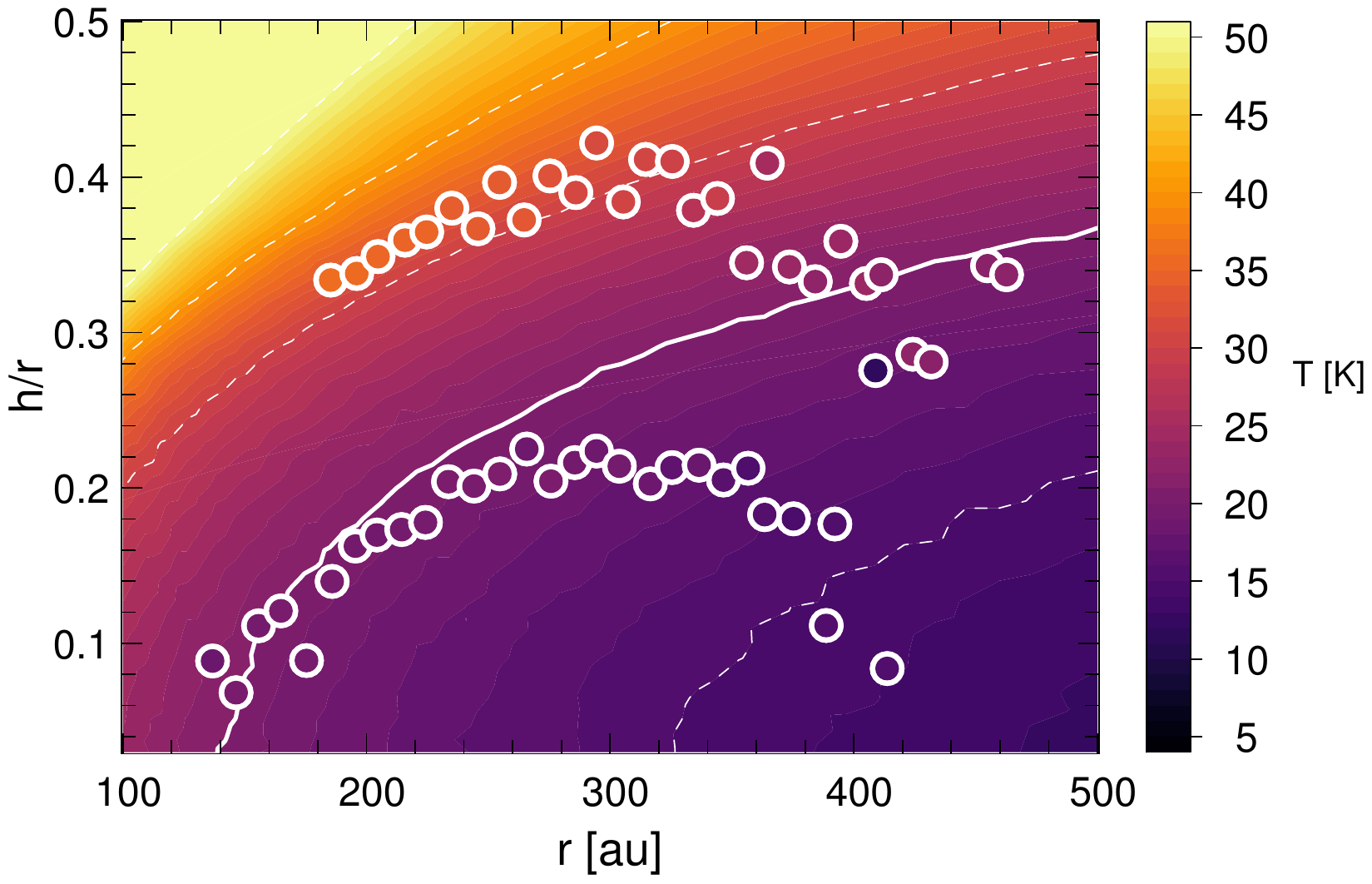}
  \caption{Comparison of the measured temperature (points)
    with the temperature structure from our MCFOST model (background). The
    same colour map is used for the data and the model. The white contour
    is the T~$=21$\,K level, and the
    dashed contours correspond to T~$=15$,
    30, 40, and 50\,K.\label{fig:T_model}}
\end{figure}

In the upper disc surface, we find that
the $^{12}$CO temperature is decreasing with radius, as expected from direct
heating by the star (Fig.~\ref{fig:h_CO}, right panel and Fig.~\ref{fig:T_model}). In the lower surface, on the other hand, the
$^{12}$CO temperature is virtually constant at $\approx$ 21\,K, between 130 and
300\,au. Because we are looking at this
surface through the disc, we are looking at the emitting CO layer  closest to the midplane.
At these large distances from the star, the
dust continuum is optically thin (see Appendix \ref{sec:cont})
and the measured CO brightness temperature is attenuated by at most 3\,\% by continuum absorption.
Because the whole CO
layer is optically thick, we are observing its top in the upper surface, and its bottom in
the lower surface.
The difference in brightness temperature between the two sides of the disc is
then a direct observation of the vertical temperature gradient.

The constant 21\,K of the lower surface is consistent with the expected CO
condensation temperature for binding onto pure CO ice \citep[{e.g.}][]{Sandford88,Bisschop06,Martin-Domenech14,Fayolle16}.
We interpret it as
a strong indication that we
are seeing gaseous CO just above the freeze-out region. This is reinforced by the measured
brightness temperature of the $^{13}$CO on the upper surface, which is extremely close
to the $^{12}$CO brightness temperature on the lower surface
(Fig.~\ref{fig:h_CO}, right panel). The combination
of $^{12}$CO and $^{13}$CO allows us to probe both sides of the gaseous CO just
above the freeze-out region. In other words, our method allows us to directly
map the vertical snow line as a function of radius. The very similar temperature profiles, despite a
typical factor $\approx 70$ in abundance, indicate that the lines remain
optically thick and that we are indeed measuring the gas temperature in this
region, hence directly measuring the CO condensation temperature.

A range of CO freeze-out temperatures have been estimated indirectly via thermo-chemical
  modelling and/or chemical imaging for the discs around the T~Tauri star TW~Hya
   (\citealp{Qi13}: 17\,K, \citealp{Zhang17}: 27\,K)
   and the Herbig~Ae star HD~163296 (\citealp{Qi11}: 19\,K, \citealp{Mathews13}: 19\,K, \citealp{Qi15}: 25\,K), where freeze-out temperatures higher than 25\,K would suggest ice
   mixtures where the binding energy is higher
   \citep{Collings03,Collings04,Noble12}.
   The most meaningful comparison is probably with the results of
   \cite{Schwarz16}, who also measured
  directly the gas kinetic temperature from the peak brightness of the CO
  isotopologue lines of TW~Hya.
   Interestingly, their measured freeze-out temperature of 21\,K is very close to
  the value we obtained for IM~Lupi. In particular, the plateau observed in the kinetic
  temperature at $\approx$ 21\,K (see their  Fig. 3a) is strikingly similar to
  the one we observe for IM~Lupi (Fig.~\ref{fig:h_CO}, right panel). This
  suggests that the ice composition is likely similar, despite the significant
  age difference: 0.5 -- 1\,Myr for IMLupi \citep{Mawet12} compared to 2--10\,Myr for TW\,Hya \citep{Barrado06,Vacca11}.

The C$^{18}$O emission has a brightness temperature of $\approx$ 8\,K on
both surfaces between 150 and 300\,au. With an excitation temperature of 21\,K
as for $^{13}$CO, this means
that the C$^{18}$O optical depth is of the order of 0.5. With a typical
abundance ratio of 8:1, the corresponding $^{13}$CO optical depth is $\approx
4$, confirming our assumption of $T_\mathrm{ex} = T_\mathrm{kin}$. This
moderate optical depth also explains why we observe the gas $^{13}$CO
in the dense region just above the CO snow line, but not at higher altitudes
where it becomes optically thin (while $^{12}$CO remains optically thick) due to the reduced densities.

At distances greater than 300\,au, where the gas density drops, the temperature
of the CO region above the snow line
appears to decrease (as measured by both $^{12}$CO and $^{13}$CO) with distance down to
$\approx$15\,K at 400\,au. This is consistent with the expected dependence of the CO
condensation temperature on the density
\citep[{e.g.}][]{Collings03,Bisschop06,Hollenbach09}. Detailed modelling
by \cite{Cleeves16} has shown that UV photo-desorption naturally explains the
diffuse CO emission outside 400\,au (see also Fig.~\ref{fig:models}d). We might
also observe this effect between 300 and 400\,au, where the increasing external
radiation field can allow the CO to remain in the gas phase at lower and lower
temperatures as the density decreases with radius.
Another possibility is that we are starting to see
deviations from local thermodynamic equilibrium as the density becomes
progressively lower than the critical density, and the measured excitation
temperature can become lower than the actual kinetic temperature.

Closer in, the emission along the isovelocity curve becomes very compact and
no longer  fills the beam, resulting in a drop in the brightness
temperature which can no longer be used to estimate the gas temperature. Interestingly, this drop does not occur at the same radius for the high
altitude $^{12}$CO emitting region ($r\approx 160\,$au) and for the CO layer
just above the snow line ($r\approx 130\,$au).
This is a direct indication that, in this range of radius, the vertical
temperature gradient is becoming steeper at high altitude in the disc, as
expected from models of irradiated discs (Fig.~\ref{fig:T_model}).
When the beam becomes larger than the actual emitting region, a fraction of the beam
is filled with emission originating from the wings of the local spectral
line, {i.e.} with lower optical depth, meaning  that part of the emission
 originates from further away along the line of sight. If the temperature gradient
is shallow, the emission will remain at the same level, as seen for $^{13}$CO, C$^{18}$O,
and $^{12}$CO on the lower disc surface.
This is consistent with the similar temperature we obtain
while observing this layer from both sides with $^{13}$CO and $^{12}$CO.

\section{Concluding remarks}

We have presented a general framework to measure directly the altitude,
velocity, and temperature of the CO emitting layers in protoplanetary discs
at intermediate inclination.
These simple geometrical considerations were applied to IM~Lupi.
Because these measurements are performed directly on the ALMA channel maps and
do not rely on any modeling of the disc, they put {unbiased} constraints on the disc
structure, which in turn are critical to feed models of disc structure, dust
evolution, and early stages of planet formation.

The results are  in agreement with the broad picture of how
we understand protoplanetary discs, with a tapered-edge disc structure passively
heated by the central star. The accuracy of our measurements is
currently limited by the spatial resolution of $\approx\,0.4$'' of our data. The presented
framework is generic. We applied it to IM Lupi which holds an unusually large
disc. However, the required high spectral resolution ($\lesssim 0.1$\,km/s)  can
be reached by ALMA down to spatial resolutions of $\approx 0.1$ -- $0.2$'' in a
few hours to detect emission with $T_\mathrm{b} \approx$ 20\,K, making it
possible to reproduce this analysis for a significant number of discs
and constrain their thermal and kinematic structures in a consistent way.

At these high spatial resolutions, discs very often display structures, such as
dust rings and gaps \citep{ALMA_HLTau,Isella16,Fedele17,van-der-Plas17}. The
presented framework may offer a way to constrain local variations in
altitude, temperature, and/or velocity of the CO layers in the
vicinity of the observed structures,
potentially pinpointing the mechanisms at the origin of these structures.

\bibliographystyle{aa.bst}
\bibliography{biblio}

\begin{acknowledgements}
We thank T.J.~Haworth for the useful discussions.
This paper makes use of the following ALMA data: ADS/JAO.ALMA\#2013.1.00798.S.
ALMA is a partnership of ESO (representing its member
states), NSF (USA), and NINS (Japan), together with NRC (Canada),  NSC and
ASIAA (Taiwan), and KASI (Republic of Korea), in cooperation with the Republic
of Chile. The Joint ALMA Observatory is operated by ESO, AUI/NRAO, and NAOJ.
The National Radio Astronomy Observatory is a facility of the National Science Foundation operated under
cooperative agreement by Associated Universities, Inc.
We acknowledge funding
from ANR of France under contract number
ANR-16-CE31-0013,
from the Australian Research Council (ARC) under
the Future Fellowship number FT170100040,
from the National Centre for Competence in Research PlanetS
supported by the Swiss National Science Foundation, and from the European Union
Seventh Framework Programme FP7-2011 under grant agreement no 284405.
\end{acknowledgements}

\begin{appendix}

  \section{C$^{18}$O observational results}
  \label{app:C18O}

\begin{figure}
    \includegraphics[width=\hsize]{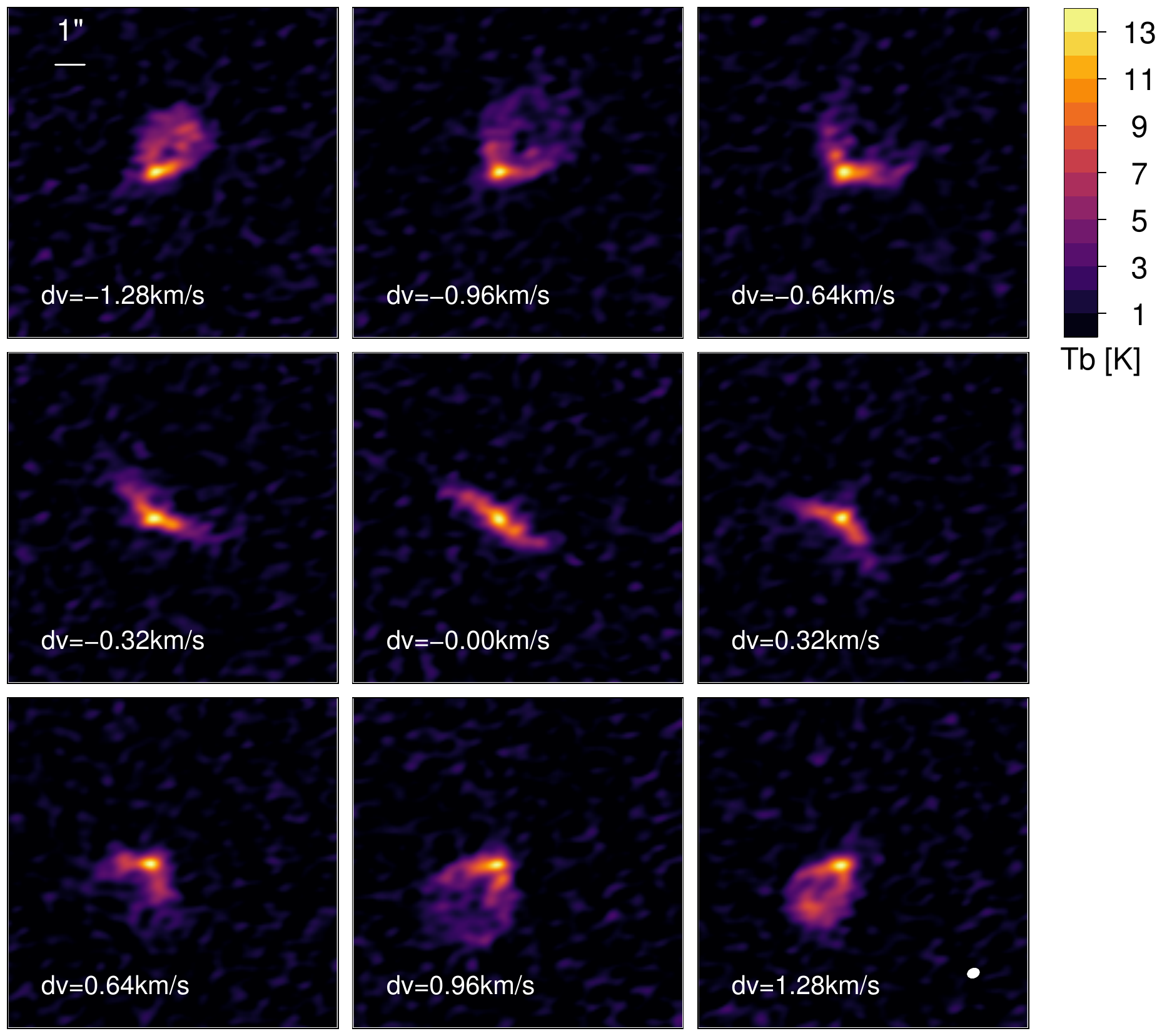}
    \caption{C$^{18}$O channel map, with the same channel
      width of 80\,m/s as in Fig.~\ref{fig:12CO}.\label{fig:C18O}}
  \end{figure}

\begin{figure}
    \includegraphics[width=\hsize]{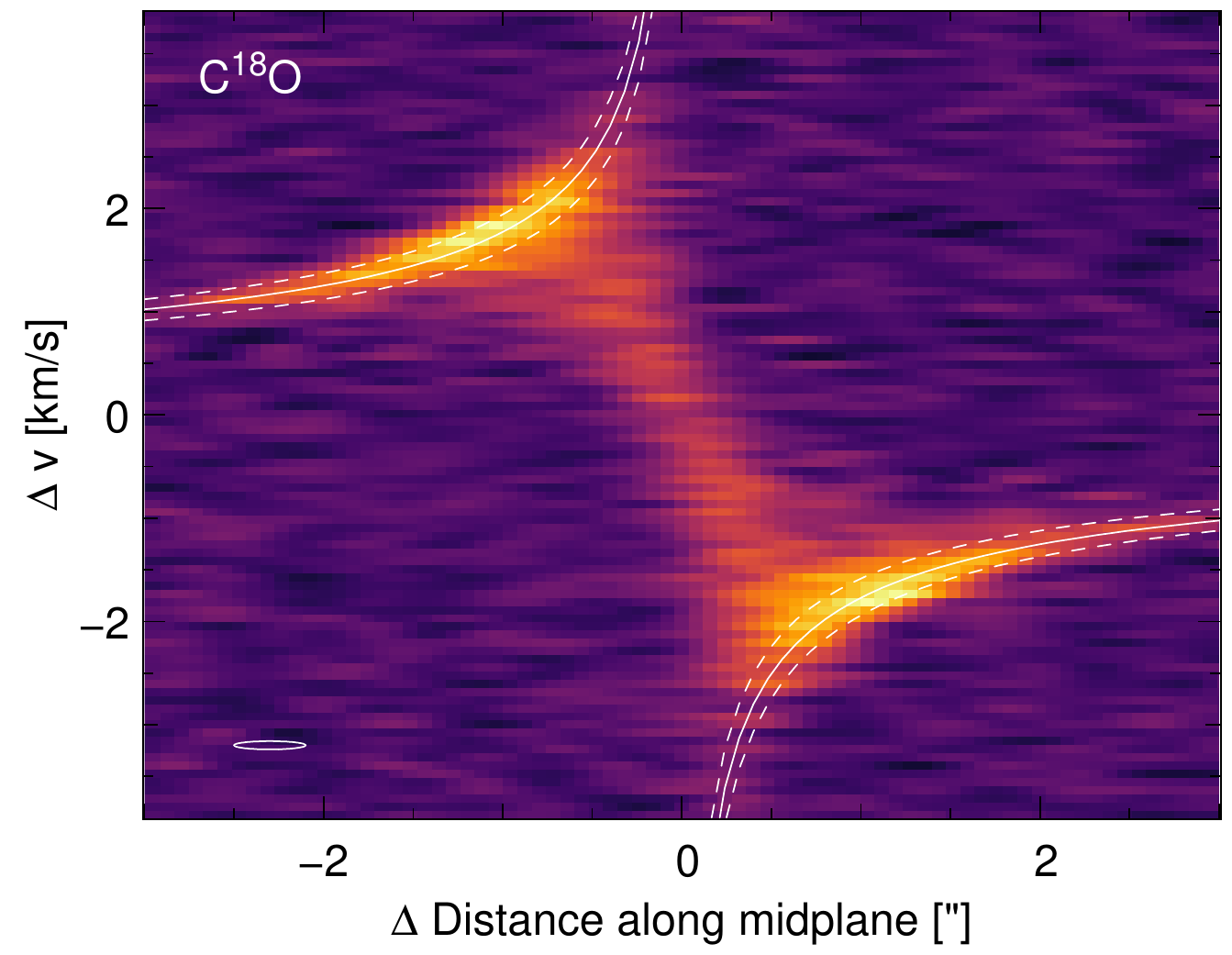}
    \caption{Position--velocity diagram of the continuum subtracted  C$^{18}$O line along the disc
      major axis. The $\Delta$v is relative to the
      systemic velocity of 4.5\,km.s$^{-1}$. The solid line
      shows the midplane Keplerian velocity for a stellar mass of 1\,M$_\odot$
      and an inclination of 48$^\circ$. The dashed lines correspond to stellar
      masses of 0.9 and 1.1\,M$_\odot$. The data have been spectrally smoothed
      to a resolution of 80\,m.s$^{-1}$. The spectro-spatial resolution is
      indicated in the lower left corner.\label{fig:PV}}
  \end{figure}

  The C$^{18}$O channel maps are presented in Fig.~\ref{fig:C18O}.
  In Figure \ref{fig:PV}, we compare position--velocity diagrams along the disc
  major axis with curves representing Keplerian
  rotation in a geometrically thin disc with
  an inclination of 48$^\circ$.
The position--velocity diagrams are nicely
reproduced by a central mass of $1\pm0.1$\,M$_\odot$, assuming a distance of
161\,pc. This is in agreement with the value used by \cite{Panic09} (rescaling
for the difference in distance) and \cite{Cleeves16}.

  \section{Model description and parameters}
  \label{app:model_parameters}

  We use the Monte Carlo radiative transfer code MCFOST \citep{Pinte06,Pinte09}
  to build a simple model of the CO emission. Our model is an extension of the
  model presented in \cite{Pinte08}, where the radial power-law surface density
  was replaced by a tapered-edge structure, following \cite{Panic09}.
  This model is only meant to provide a qualitative description of the
  ALMA band 6 data. A complete fit of all the observations of IM Lupi is
  beyond the scope of this paper, and we refer the readers to the work by
  \cite{Cleeves16} for the most recent global model of IM Lupi.

  The central object is described by a photosphere at 3\,900\,K, a radius of
  2.15\,R$_\odot$ and a mass of 1\,M$_\odot$. An ultra-violet excess is added to
  the phosphere, defined by a parameter $f_\mathrm{UV}$, which describes the
  fractional luminosity excess between 91.2 and 250\,nm.

  An external, isotropic, interstellar radiation field is
    added in model $d$ presented in Fig.~\ref{fig:models}.
    It is composed of a highly diluted UV field, an infrared background
    radiation field, and the cosmic microwave background.
    We  follow the same prescription as \citet[][see their Appendix
    A.4 for details]{Woitke16}. In this work, the UV component is scaled to
    4\,G$_\circ$, following the results of \cite{Cleeves16}, who estimated the
    local radiation field from geometrical arguments based on \mbox{HIPPARCOS} data.

  We consider an axisymmetric flared density structure with a Gaussian vertical
  profile. The disc scale height is described by a power-law
  \begin{equation}
    h(r) = h_0 (r/r_0)^\beta
  \end{equation}
  and its surface density by
  \begin{equation}
    \Sigma(r) \propto r^{-\gamma} \exp\left( - (r/r_\mathrm{tap})^{2-\gamma_{exp}} \right),
  \end{equation}
  where r is the radial coordinate in the equatorial plane and $h_0$ is the
  scale  height at the radius $r_0 = 100$\,au.
  The disc extends from $r_\mathrm{in}$, that
  we fixed to 0.3\,au, to an outer $r_\mathrm{out}$, set to 8 times the
  critical radius $r_\mathrm{tap}$, {i.e.} where the density is so low
  that the actual value does not affect any of the synthetic observables.

  Dust grains are defined as homogeneous and spherical particles
  with sizes distributed according to the power law $\mathrm{d}n(a) \propto
  a^{-3.5} \mathrm{d}a$ between $a_\mathrm{min} = 0.03\,\mu$m and
  a$_\mathrm{max}$. Optical properties are computed using the Mie theory,
  adopting the same olivine stoichiometry from \cite{Dorschner95} as in
  \cite{Pinte08}.

  Vertical dust settling is implemented by following the same prescription as
  in \cite{Pinte16} and \citet[see their equation 19]{Fromang09}, where the
  degree of dust settling is set by varying the turbulent viscosity
  coefficient $\alpha_\mathrm{SS}$ \citep{Shakura73}.

  For the dust radial migration, we follow the simple prescription introduced in
  \cite{Pinte14}. The maximum
  expected grain size is  set to the optimal size of migration
  \begin{equation}
    s_{\mathrm{opt}}(r) = \frac{\rho(r,z=0)\, c_{\mathrm{s}}(r,z=0)}{\rho_\mathrm{dust}\, \Omega_{\mathrm{K}}} ,
    \label{eq:sopt}
  \end{equation}
  where $\rho_{\mathrm{g}}$ and $c_{\mathrm{s}}$ are the gas density and
  sound speed, $\rho_\mathrm{dust}$ the intrinsic dust density, and
  $\Omega_{\mathrm{K}}$ the   Keplerian frequency respectively.

  CO chemistry is dominated by a few key mechanisms:  photo-dissociation by UV
  radiation in the upper layers,  condensation on the dust grains at
  low temperatures (T $\lesssim$ 20\,K), and  photo-desorption of the CO ice
  by UV radiation in the outer disc.
  We have implemented a very simple model of the CO chemistry, which was
  calibrated from ProDiMo models \citep[][see e.g. their
  Fig. C.2]{Woitke16}. We initially assume that the CO abundance $X$(CO) is
  constant everywhere.  CO is photo-dissociated, {i.e.} we set its gas
  abundance to 0  if
  \begin{equation}
    \log (\chi/n) > -6
  ,\end{equation}
  where $n$ is the number density of hydrogen atoms and
  $\chi$ describes the intensity of the UV radiation field, following \cite{Woitke09}
  \begin{equation}
    \chi = \frac{1}{F^\mathrm{Draine}}\int_{91.2\,\mathrm{nm}}^{205\,\mathrm{nm}}
    u_\lambda\,\mathrm{d}\lambda
  ,\end{equation}
  with F$^\mathrm{Draine} = 1.921\times10^{12}\,\mathrm{m^{-2}\,s^{-1}}$, and where
  $u_\lambda$ is the energy density of the radiation field which was computed
  by MCFOST at each point in the model.

  CO is frozen out on the dust grains if $T < 21K$, {i.e.} its abundance is
  multiplied by a factor $\epsilon < 1$, except if
  \begin{equation}
   \log (\chi/n) > -7,
 \end{equation}
 where CO is photo-desorbed.

 The dust temperature structure is computed using MCFOST.
 We then assume that the gas temperature is equal to the dust
 temperature and that the CO emission is at local thermodynamic equilibrium,
 and compute the synthetic channel map via a ray-tracing method.

 As our aim was simply to build an illustrative model, we only adjusted the
 Band 6 CO data in the image place via a genetic algorithm (each generation was
 composed of 100 models and the genetic algorithm was run for 50
 generations). As a consequence, our model may not be a unique
 solution, and the model parameters presented in
 Table~\ref{tab:model_parameters} must be interpreted with care.

\begin{table}
  \begin{tabular}{lc}
    \hline
    \hline
    Model parameter & value\\
    \hline

    $a_\mathrm{max}$ [$\mu$m]& 2300\\
    M$_\mathrm{dust}$ [M$_\odot$] & $1.75 \times 10^{-3}$ \\
    gas/dust radio & 347 \\
    $\alpha_\mathrm{SS}$ (dust settling) & $9 \times10^{-3}$\\
    $\beta$  & 1.19\\
    $\gamma$  & 0.48 \\
    $\gamma_\mathrm{exp}$  & 0.32 \\
      R$_\mathrm{tap}$ [au] & 284\\
      h at 100\,au [au] & 12.9 \\
      $\varv_\mathrm{turb}$ [km/s] & 0.08 \\
      $X$($^{12}$CO) & 5$\times 10^{-5}$\\
      $X$($^{13}$CO) & 7$\times 10^{-7}$\\
      $X$(C$^{18}$O) & 1$\times 10^{-7}$\\
      $\epsilon$ & 8$\times 10^{-5}$\\
      $f_\mathrm{UV}$ & 0.09\\
    \hline
  \end{tabular}
  \caption{Parameters of our disc model, as derived by our genetic algorithm
    from fitting of the Band 6 ALMA data only.\label{tab:model_parameters}}
\end{table}

\section{Continuum optical depth}
\label{sec:cont}

\begin{figure}
  \centering
\includegraphics[width=0.8\hsize]{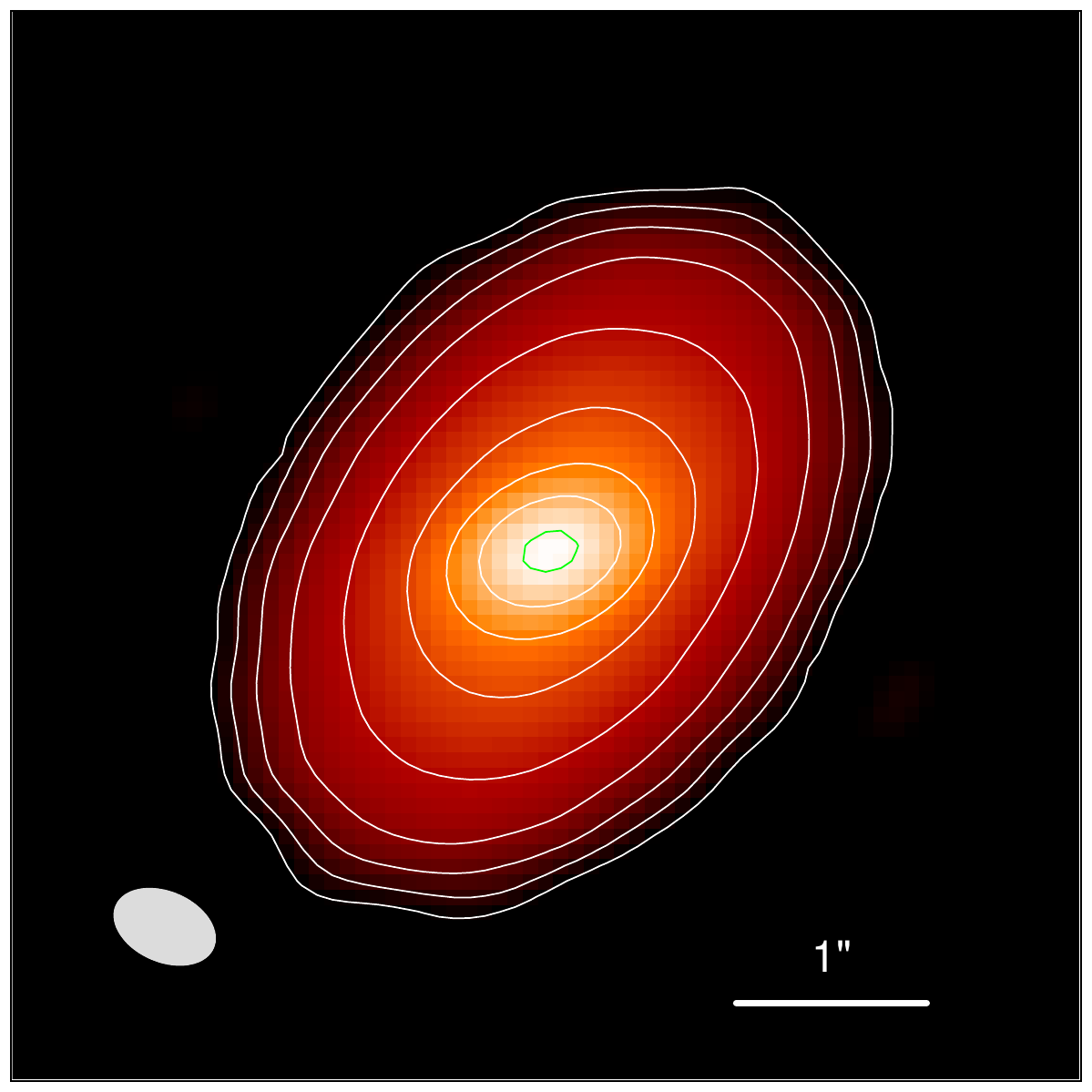}
\caption{Continuum image of IM Lupi at $\lambda=1.3\,$mm.
Contours begin at 4\,$\sigma$ and step in factors of 2 in intensity up to
1024\,$\sigma$ (green contour),  where $\sigma=0.05$\,mJy/beam is the RMS measured on the map
away from the source.
\label{fig:continuum_image}}

\end{figure}

\begin{figure}
\includegraphics[width=\hsize]{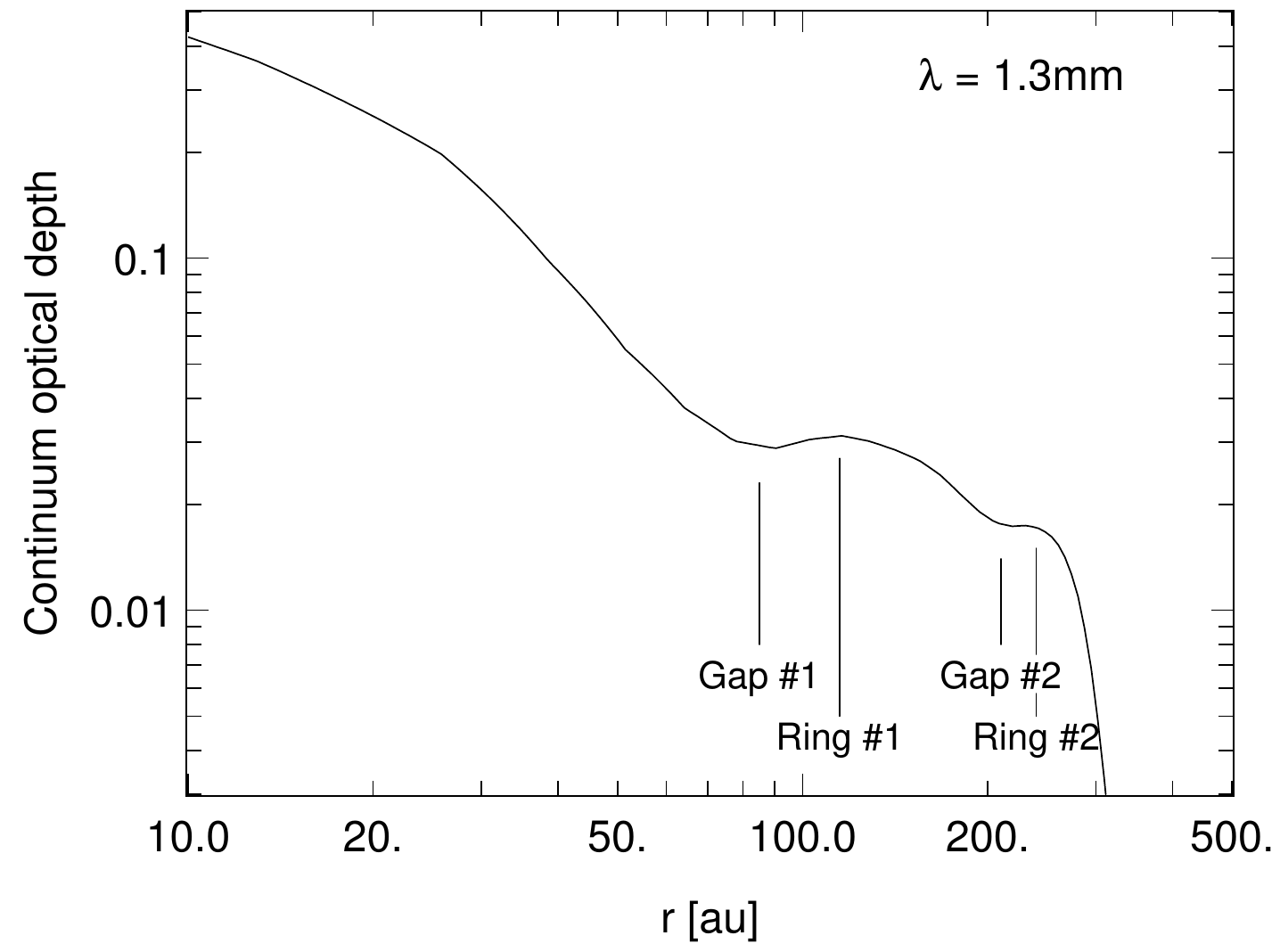}
\caption{Continuum  optical depth along the line of sight at $\lambda=1.3\,$mm as a function
  of radius. The optical depth takes the inclination of the disc into account
  and is measured along the disc major axis.
  \label{fig:cont_optical_depth}}
\end{figure}

To assess the continuum optical depth of the disc, a more accurate radiative
transfer model of the continuum emission (Fig.~\ref{fig:continuum_image}) is
required.
We use the same iterative procedure as
in \cite{Pinte16} for HL Tau.
In short, we first extract a brightness profile along the disc major axis
in the CLEANed map (average of both sides). We
deconvolve this profile by a Gaussian corresponding to the size of the beam in that direction.
We convert the deconvolved brightness profile into a surface density profile
assuming an initial power-law radial temperature profile (exponent = -0.5),
from which we compute a MCFOST model. We compare the resulting synthetic surface
brightness profile with the observed one and iteratively correct the surface
density until the synthetic brightness profile does not change by more than 1\%
at any radius. We reach convergence after about ten iterations.
The resulting continuum optical depth as a
function of radius is presented in Fig.~\ref{fig:cont_optical_depth}.

In the regions we are studying in this paper, {i.e.} outside  100\,au,
the maximum optical depth is $\approx 0.03$, meaning that the flux on the
lower surface of the disc is attenuated by at most 3\,\%, due to continuum absorption by the dust.

Interestingly, the optical depth and corresponding dust surface density show two dips at
$\approx 85$ and 210\,au, suggesting the presence of two gaps, and the
corresponding two rings peaking at $\approx 115$ and 240\,au.
They very likely correspond to the rings observed by \cite{Cleeves16} at
$\approx$150 and 250\,au, after applying an unsharp mask.

\end{appendix}

\end{document}